\documentclass[11pt,twoside]{article}

\usepackage{asp2006}
\usepackage{epsf}
\usepackage{lscape}
\usepackage{graphicx}

\begin{document}
\title{How Common are Extrasolar, Late Heavy Bombardments?}   
\author{Mark Booth,\altaffilmark{1} Mark C. Wyatt,\altaffilmark{1} Alessandro Morbidelli,\altaffilmark{2} Amaya Moro-Mart{\'{\i}}n\altaffilmark{3,4} and Harold F. Levison\altaffilmark{5}}   
\altaffiltext{1}{Institute of Astronomy, Madingley Rd, Cambridge CB3 0HA, UK}
\altaffiltext{2}{Observatoire de la C\^ote d'Azur, Nice, France}
\altaffiltext{3}{Centro de Astrobiologia - CSIC/INTA, 28850 Torrej\'on de Ardoz, Madrid, Spain} 
\altaffiltext{4}{Department of Astrophysical Sciences, Peyton Hall, Ivy Lane, Princeton University, Princeton, NJ 08544, USA}
\altaffiltext{5}{Department of Space Studies, Southwest Research Institute, Boulder, CO 80302, USA}    

\begin{abstract} 
The habitability of planets is strongly affected by impacts from comets and asteroids. Indications from the ages of Moon rocks suggest that the inner Solar System experienced an increased rate of impacts roughly 3.8 Gya known as the Late Heavy Bombardment (LHB). Here we develop a model of how the Solar System would have appeared to a distant observer during its history based on the Nice model of Gomes et al. (2005). We compare our results with observed debris discs. We show that the Solar System would have been amongst the brightest of these systems before the LHB. Comparison with the statistics of debris disc evolution shows that such heavy bombardment events must be rare occurring around less than 12\% of Sun-like stars.
\end{abstract}


%
%
As well as the evidence for the LHB on the Moon\citep{mbtera74}, there is evidence that the rest of the Solar System also went through a period of increased bombardment including the Earth \citep{mbkring02, mbjorgensen09}. Impacts on the Earth from comets and asteroids may have caused mass extinctions like that which may have wiped out the dinosaurs. However, the earliest evidence for life on Earth comes from just after the time of the LHB \citep{mbmojzsis96}. Since large impacts have the potential to cause the formation of deep sea vents, which may be the sites where life first developed, it is plausible that the LHB was, in fact, necessary for life \citep{mbkring03}. 

Whether the LHB helped or hindered life, it almost certainly had an effect on life and so it is interesting to see whether other planetary systems would also have gone through a late phase of heavy bombardment like the Solar System. To answer this question, we need to know what observable properties of the Solar System would change during and after an LHB event. 

Planetesimals in any planetary systems are constantly undergoing collisional evolution. This produces dust and can be observed as a debris disc. \citet{mbgomes05} model the dynamics of planetesimals during the LHB. By making some assumptions about the planetesimals and their size distribution, we use the evolution of the mass distribution from this model to find out how the flux from the Kuiper belt changes due to the LHB \citep{mbbooth09}.

\begin{figure}[tb]
	\centering
		\includegraphics[clip=true,width=\columnwidth]{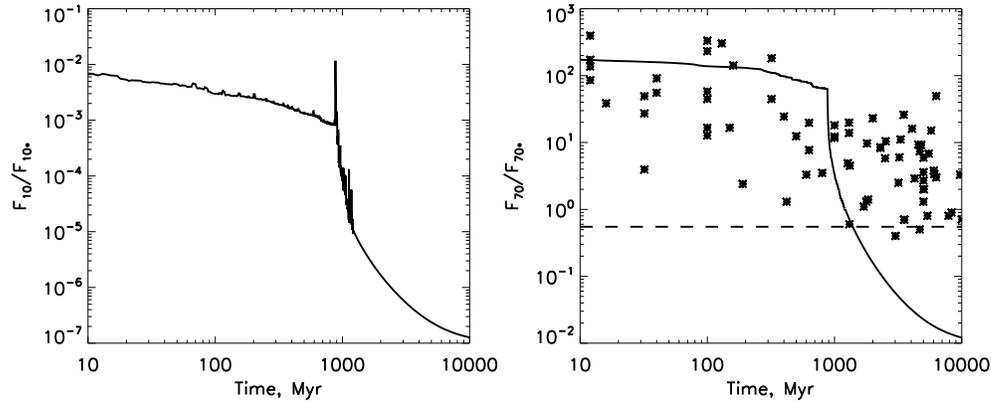}
		\caption{Excess ratio versus time for 10~$\mu$m (left) and 70~$\mu$m (right). The solid line represents the emission from our model. The asterisks are observed discs and the dashed line shows the approximate observational limit.}
	\label{f2470evol}
\end{figure}

Figure \ref{f2470evol} shows the emission from the Solar System at 10 and 70~$\micron$. The figure also shows observed debris discs from the literature. This shows that the Solar System would have been amongst the brightest discs before the LHB and that its emission rapidly dropped after the LHB. There is also a rise in mid-IR emission during the LHB, which could explain some observed debris discs that have mid-IR excesses above what would be expected from steady state evolution of debris discs \citep{mbwyatt07}. \citet{mbtrilling08} survey debris discs and find that 16\% of stars are observed with 70~$\micron$ and that this number remains approximately constant with age. By assuming that any star that is observable at 70~$\micron$ when it is young and not observable when it is old has undergone an LHB-like event, we can constrain the number of stars that would have undergone an LHB to 12\%.

\acknowledgements 
MB acknowledges the financial support of a PPARC / STFC studentship.


\end{document}